\documentclass[prb,twocolumn,superscriptaddress,groupedaddress,amsmath,amssymb]{revtex4-2}
\usepackage[utf8]{inputenc}
\usepackage[english]{babel}

\usepackage{color}
\usepackage{graphicx}
\usepackage{bm}
\usepackage{xcolor}
\usepackage{epstopdf}
\usepackage{amsmath}
\usepackage{amssymb}
\usepackage{verbatim}

\usepackage{hyperref}
\hypersetup{
	breaklinks=true,   
	colorlinks=true,
	urlcolor=blue,
	pdfusetitle=true, 
}

\newcommand{\bs}{\boldsymbol}
\newcommand{\bpm}{\begin{pmatrix}}
\newcommand{\epm}{\end{pmatrix}}

\newcommand{\sig}{\mbox{\boldmath{$\sigma$}}}
\newcommand{\tauu}{\mbox{\boldmath{$\tau$}}}
\definecolor{purpleheart}{rgb}{0.41, 0.21, 0.61}

\begin{document}

\title{RKKY interaction at helical edges of topological superconductors}

\author{Katharina Laubscher}
\affiliation{Department of Physics, University of Basel, Klingelbergstrasse 82, CH-4056 Basel, Switzerland}
\author{Dmitry Miserev}
\affiliation{Department of Physics, University of Basel, Klingelbergstrasse 82, CH-4056 Basel, Switzerland}
\author{Vardan Kaladzhyan}
\affiliation{Department of Physics, University of Basel, Klingelbergstrasse 82, CH-4056 Basel, Switzerland}
\author{Daniel Loss}
\affiliation{Department of Physics, University of Basel, Klingelbergstrasse 82, CH-4056 Basel, Switzerland}
\author{Jelena Klinovaja}
\affiliation{Department of Physics, University of Basel, Klingelbergstrasse 82, CH-4056 Basel, Switzerland}

\date{\today}

\begin{abstract}
We study spin configurations of classical magnetic impurities placed close to the edge of a two-dimensional topological superconductor
both analytically and numerically. First, we demonstrate that the spin of a single magnetic impurity close to the edge of a topological superconductor tends to align along the edge. The strong easy-axis spin anisotropy behind this effect originates from the interaction between the impurity and the gapless helical Majorana edge states. We then compute the Ruderman-Kittel-Kasuya-Yosida (RKKY) interaction between two magnetic impurities placed close to the edge.
We show that, in the limit of large interimpurity distances, the RKKY interaction between the two impurities is mainly mediated by the Majorana edge states and leads to a ferromagnetic alignment of both spins along the edge. This effect could be used to detect helical Majorana edge states.
\end{abstract}

\maketitle

\section{Introduction}

Two magnetic impurities placed on a host material can effectively interact by coupling to the electron spin density of the host. This so-called Ruderman-Kittel-Kasuya-Yosida (RKKY) interaction~\cite{Ruderman1954,Kasuya1956,Yosida1957} is crucial in determining the magnetic ordering of the impurities and has recently come into focus due to its key role in designing magnetic-impurity-based \textit{ad hoc} topological superconductors (TSCs) hosting Majorana zero modes~\cite{Nadj-Perge2013,Pientka2013,Braunecker2013,Klinovaja2013a,Vazifeh2013,Pientka2014,Poyhonen2014,
	Reis2014,Kim2014,Li2014,Heimes2014,Brydon2015,Weststrom2015,Peng2015,Hui2015,Rontynen2015,
	Braunecker2015,Hsu2015,Poyhonen2016,Zhang2016,Li2016a,Rontynen2016,Hoffman2016,Li2016b,Schecter2016,
	Christensen2016,Kaladzhyan2017a,Andolina2017,Kobialka2020,Menard2017,Menard2019,Pawlak2016,Nadj-Perge2014,Ruby2015,Ruby2017,Feldman2016,Kim2018,Pawlak20191,Ding2021,Laubscher2021}. The exact form of the RKKY interaction depends on the properties of the underlying host material and has been extensively studied for various bulk systems \cite{Zyuzin1986,Abrikosov1988,Demokritov1992,Poilblanc1994,Balatsky1995,Aristov1997,
	Galitski2002,Hindmarch2003,Imamura2004,Saremi2007,Hwang2008,Simon2008,Braunecker2009,Braunecker2009prl,Schulz2009,Black-Schaffer2010,Braunecker2010,Chesi2010,Sherafati2011,Kogan2011,Tokura2012,Klinovaja2013b,Parhizgar2013,Yao2014,Hoffman2015,Schecter2015,Tsvelik2017,Yevtushenko2018}. Generally, the RKKY interaction in metals decays as a power law in the interimpurity distance with an oscillatory prefactor, while it is exponentially suppressed in insulators and superconductors.

More recently, it was realized that the presence of boundaries can lead to interesting modifications to the RKKY interaction. Such boundary effects were studied, for example, in topologically trivial $s$- and $d$-wave superconductors~\cite{Deb2021,Ghanbari2021}. Furthermore, since boundary effects are expected to be particularly interesting in topological materials, several works have studied magnetic impurities coupled to edge or surface states of topological insulators (TIs)~\cite{Liu2009,Biswas2010,Garate2010,Zhu2011,Abanin2011,Zyuzin2014,Efimkin2014,Gao2009,Meng2014,Lee2015,Yang2016,KKBurmistrov2017,Hsu2017,Hsu2018}. For TSCs, on the other hand, only a few studies focusing on quantum spins coupled to 1D Majorana edge states exist~\cite{Shindou2010,Zitko2011,Eriksson2015}. So far, these rely on effective 1D models for the edge states without explicitly describing the full 2D TSC and its boundary.

\begin{figure}[tb]
	\centering
	\includegraphics[width=1\columnwidth]{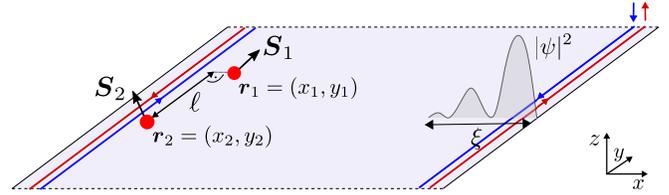}
	\caption{A 2D $p$-wave TSC hosts helical Majorana edge states shown in red and blue. Their wave functions  $|\psi|^2$  decay exponentially into the bulk on the scale of the superconducting coherence length $\xi$. Two magnetic impurities (red dots) with spins ${\bs S}_1$  and ${\bs S}_2$ (black arrows) are placed in the vicinity of the edge. Their separation along the direction of the edge is denoted by $\ell$.
	}
	\label{fig:SystemSketch}
\end{figure}

In this work, we extend and deepen the understanding of RKKY effects in topological materials by studying classical magnetic impurities close to the edge of a 2D $p$-wave TSC with helical Majorana edge states, see Fig.~\ref{fig:SystemSketch}. Here, we carefully model the full 2D system, allowing us to describe not only the behavior close to the edge but also the crossover to the bulk regime at distances from the edge larger than the edge state localization length. Our main findings for impurities close to the edge can be summarized as follows. First, we find that the spin of a single magnetic impurity tends to align along the edge due to a strong easy-axis anisotropy (EAA) imposed by the symmetries of the TSC Hamiltonian. Second, two magnetic impurities separated by large distances along the edge interact mainly through the Majorana edge states. The corresponding RKKY interaction is of the Ising type, decays inversely proportional to the interimpurity distance, and results in a ferromagnetic alignment of the impurity spins along the edge. All of these results are derived analytically using a continuum model and independently verified by numerical exact diagonalization of the corresponding tight-binding model.

The above features stand in stark contrast to what is observed in a trivial superconductor without edge states, where the RKKY coupling is exponentially suppressed with the interimpurity distance even for impurities close to the edge. Therefore, we believe that spectroscopy of magnetic impurities~\cite{Stano2013,Nadj-Perge2014,Bergmann2015,Xu2015,Ruby2017,Cornils2017,Jeon2017,Li2018,Schneider2019,Choi2019,Jack2019,Jack2021,Wang2021} can serve as a powerful tool to experimentally probe TSCs and systems with topological edge states in general.

\section{Model}

We consider a 2D helical $p$-wave superconductor \cite{BernevigHughes2013}
described by the mean-field Hamiltonian
\begin{eqnarray}
&& \mathcal{H}(\bm k) = \frac{k^2 - k_F^2}{2 m} \tau_z + \alpha \tau_x \left(\sigma_x k_y - \sigma_y k_x\right)  \label{ham}
\end{eqnarray}
written in the Nambu basis $(\psi_{\bm k,\uparrow},\psi_{\bm k,\downarrow},\psi_{-\bm k,\downarrow}^\dagger,-\psi_{-\bm k,\uparrow}^\dagger)$, where $\psi_{\bm k,\sigma}^{(\dagger)}$ is the electron 
annihilation (creation) operator with spin $\sigma\in\{\uparrow,\downarrow\}$ and in-plane momentum $\bm k = (k_x, k_y)$,
$k_F$ is the Fermi momentum, 
$k^2 = k_x^2 + k_y^2$,
$m$ is the effective mass,
$\sig$ ($\tauu$) are the Pauli matrices
acting on the spin (particle-hole) degree of freedom, and
$\alpha>0$ is the TSC coupling constant responsible for opening the superconducting gap~\footnote{The case $\alpha < 0$ is connected to $\alpha > 0$ through the unitary transformation $H(-\alpha) = \sigma_z H(\alpha) \sigma_z$. This unitary transformation does not affect any of the presented results.}.
We set $\hbar=1$ throughout this work. 
The bulk spectrum of $\mathcal{H}$ is fully gapped:
\begin{equation}
\varepsilon_\pm (\bm k) = \pm \sqrt{\left(\frac{k^2 - k_F^2}{2 m} + m \alpha^2\right)^2 + \Delta^2} , \label{Delta}
\end{equation}
where $\Delta=\alpha \sqrt{k_F^2 - (m \alpha)^2} \approx \alpha k_F $ is the bulk superconducting gap. Explicitly, we have $\varepsilon_\pm (\bm k_0)=\pm\Delta$ for $|\bm k_0|=\sqrt{k_F^2-2m^2\alpha^2}$. We work in the regime where $\Delta \ll \varepsilon_F:=k_F^2/(2 m)$, so $k_F \gg m \alpha$.

As the mean-field Hamiltonian given in Eq.~(\ref{ham}) is rotationally invariant, all edges are equivalent. In the following, we focus on the straight edge at $x = 0$, such that the TSC occupies the half-space $x > 0$. With the boundary at $x = 0$, the system is still translationally invariant along the $y$ axis, so the momentum $k_y$ is a good quantum number. Moreover, $\mathcal{H}$ commutes with $\sigma_z \tau_z$, which allows for the following choice of eigenstates:
\begin{equation}
 \Psi_{\eta, k_y, n} (x, y) = e^{i k_y y} \Psi_{\eta, k_y, n} (x) , \label{ky} \\
\end{equation}
where $\eta = \pm 1$ denotes the eigenvalue of $\sigma_z \tau_z$ [i.e., $\sigma_z \tau_z \Psi_{\eta, k_y, n} (x) = \eta \Psi_{\eta, k_y, n} (x)$], $k_y$ is the momentum along the $y$ axis,
and $n$ labels all other quantum numbers in the system. 
The boundary at $x = 0$ obliges all quantum states to satisfy the boundary condition $\Psi_{\eta, k_y, n} (x = 0) = 0.$

The topologically nontrivial nature of the TSC is reflected in the presence of Majorana edge states, see Fig.~\ref{fig:SystemSketch}. Their wave functions can be found analytically as
\begin{equation}
 \Psi_{\eta, k_y} (x) = \sqrt{\frac{2}{\xi} \left(k_F^2 - k_y^2 \right)} \frac{\sin (\kappa x)}{\kappa} e^{-\frac{x}{\xi}} {u_\eta \choose -\eta u_{-\eta}}, \label{edgegap} \\
\end{equation}
where we have defined $\kappa = \sqrt{k_F^2 - (m \alpha)^2 - k_y^2}$, $u_+^T = (1,0)$, $u_-^T = (0, 1)$, and where the localization length $\xi = 1/(m \alpha) \approx v_F/\Delta$ defines the effective width of the edge states. The above edge state wave functions are normalizable for all $k_y \in (-k_F, k_F)$ since $1/\xi\pm\mathrm{Im}(\kappa)>0$ for these values of $k_y$. The corresponding eigenenergies are given by $\varepsilon_{\eta, k_y} = - \eta \alpha k_y$. 

In the following, we place magnetic impurities in the vicinity of the TSC edge. The impurity spins interact with the itinerant electrons via the local exchange interaction
\begin{equation}
 \mathcal{H}_\mathrm{imp} = J \sum_i \sig \cdot \bm S_i \, \delta(\bm r - \bm r_i) , \label{spinspin}
\end{equation}
where $\bm S_i$ is the spin of the magnetic impurity placed at the position $\bm r_i =(x_i,y_i)$ with $x_i>0$. The impurity spins are considered large, $S=|\bm S_i| \gg 1$, which allows us to treat them as classical vectors. This regime is expected to be relevant, e.g., for transition metal adatoms with spins $S\geq 3/2$~\cite{Nadj-Perge2014,Ruby2017,Cornils2017,Jeon2017,Li2018,Schneider2019,Ruby2015,Ji2008,Ruby2015b,Ruby2016,Ruby2018,Choi2018,Choi2018b}, where quantum effects are expected to be small and theoretical predictions based on classical spins have reasonably explained experimental findings in the past. In particular, the classical approximation allows us to neglect the Kondo effect since the Kondo temperature becomes exponentially small for $S\gg 1$. Additionally, the exchange coupling constant $J$ is assumed to be small compared to the electron bandwidth, $J S m  \ll 1$, so we can neglect the renormalization of the superconducting order parameter~\cite{Flatte1997a,Salkola1997,Flatte2000,Balatsky2006,Morr2006,Meng2015,Bjornson2017} close to the impurity. Appendix~\ref{app:ldos} shows that, in this regime, the local density of states is changed only slightly in the vicinity of an impurity placed close to the TSC edge. Furthermore, an impurity close to the edge of the TSC does not lead to the emergence of a Shiba bound state, see again Appendix~\ref{app:ldos}.

\begin{figure*}[bt]
	\centering
	\includegraphics[width=0.9\textwidth]{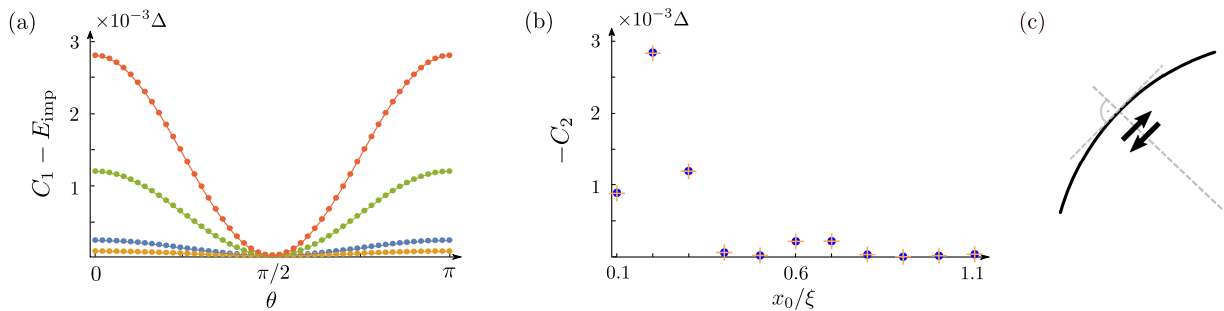}
	\caption{(a) The negative impurity-induced contribution $C_1-E_{\mathrm{imp}}$ as a function of the polar angle $\theta$ for a single magnetic impurity at a distance $x_0=0.2\,\xi$ (red), $x_0=0.3\,\xi$ (green), $x_0=0.4\,\xi$ (orange), and $x_0=0.6\,\xi$ (blue) from the edge. The dots represent numerical data obtained via exact numerical diagonalization of a discretized 2D model (see Appendix~\ref{app:tightbinding} for details) and the solid line is a fit $\propto\cos^2\theta$. We find that the total energy is minimized if the impurity spin is aligned along the edge, i.e., for $\theta=0$ and $\theta=\pi$. (b) Dependence of $C_2$ on the distance from the edge obtained via exact numerical diagonalization of a 2D system (blue circles) and via numerical evaluation of Eq.~(\ref{B2}) in a ribbon geometry (orange crosses). The two curves coincide very well and show that $C_2$ decays exponentially with the distance from the edge on the scale of the edge state localization length $\xi$. Furthermore, $C_2$ exhibits oscillations with a period of $\pi/k_F$. (c) While we focused on a straight edge for simplicity, our results can easily be generalized to more complicated edge geometries. In this case, the impurity spin prefers to align tangentially to the edge as is indicated by the black arrows. The parameters used in panels (a) and (b) are $\Delta/\varepsilon_F=0.28$ and $JSm=0.1$.}
	\label{fig:SingleImpurityTotalEnergy}
\end{figure*}

\section{Single magnetic impurity}

First, we consider a single magnetic impurity placed at the position $\bm r_0$. Since we work in the limit of weak exchange interaction, the impurity-induced correction to the total energy can be obtained perturbatively. The first-order correction vanishes in time-reversal invariant systems as there is no intrinsic magnetization in the system.
The second-order correction comes from the local magnetization induced by the impurity itself:
\begin{equation}
E_{\mathrm{imp}} = \frac{J^2}{2} \int\limits_{-\infty}^\infty \frac{d \omega}{2  \pi} {\rm Tr} \left\{\left[\sig \cdot \bm S \, G(\bm r_0, \bm r_0, i \omega) \right]^2 \right\} . \label{secondorder}
\end{equation}
Here, $G(\bm r, \bm r', i \omega)$
is the Matsubara Green function of the TSC without impurities:
\begin{eqnarray}
&& \hspace{-25pt}G(\bm r, \bm r', i\omega) = \sum_{\eta, k_y, n} \frac{\Psi_{\eta, k_y, n}(x) \Psi^\dagger_{\eta, k_y, n} (x') e^{i k_y (y - y')}}{i \omega - \varepsilon_{\eta, k_y, n}}, \label{eq:green}
\end{eqnarray}
where $\varepsilon_{\eta, k_y, n}$ is the eigenenergy of the state $\Psi_{\eta, k_y, n} (x)$
[see Eq.~(\ref{ky})].

The Hamiltonian $\mathcal{H}$ given in Eq.~(\ref{ham}) with our choice of boundary conditions is time-reversal symmetric, particle-hole symmetric, and invariant under inversion of the $y$ axis. In Appendix~\ref{app:eaa}, we show that these symmetries lead to the following  matrix decomposition for $G(\bm{r}_0,\bm{r}_0, i \omega)\equiv G(x_0, i \omega)$:
\begin{equation}
G(x_0, i\omega) = i A_1(x_0, i \omega) + A_2 (x_0, i \omega) \tau_z + i B (x_0, i \omega) \sigma_y \tau_y, \label{eq:Gmat}
\end{equation}
where $A_{1} (x, i \omega)$ and $B(x, i \omega)$ [$A_{2} (x, i \omega)$] are real-valued odd [even] functions of $\omega$. Evaluating the spin trace in Eq.~(\ref{secondorder}) using the matrix decomposition Eq.~(\ref{eq:Gmat}) [see again Appendix~\ref{app:eaa}], we find that $A_{1,2}(x, i \omega)$ give fully isotropic contributions, while $B(x, i \omega)$ additionally results in an anisotropic term:
\begin{equation}
 E_{\mathrm{imp}} = - (2 J S_y)^2 \int\limits_{-\infty}^\infty \frac{d \omega}{2 \pi} B^2(x_0, i \omega) +C_1 , \label{easy}
\end{equation}
where $C_1\propto S^2$ is the isotropic contribution. From the matrix decomposition Eq.~(\ref{eq:Gmat}), we further find that $B (x_0, i \omega)$ can be represented as follows:
\begin{eqnarray}
&& \hspace{-30pt} B (x_0, i \omega) = - \frac{\omega}{2} \sum\limits_{\eta, k_y, n}^{\varepsilon > 0} \frac{\Psi^\dagger_{\eta, k_y, n}(x_0) \sigma_y \tau_y \Psi_{\eta, k_y, n} (x_0)}{\omega^2 + \varepsilon_{\eta, k_y, n}^2} , \label{B2}
\end{eqnarray}
where the $\Psi_{\eta, k_y, n}(x)$ enlist all eigenstates of $\mathcal{H}$.
The upper limit $\varepsilon > 0$ indicates that 
only the quantum states with positive energies
$\varepsilon_{\eta, k_y, n} > 0$ are taken into account. The contribution of eigenstates with negative energies is already included in Eq.~(\ref{B2}) via the particle-hole symmetry.

From the above, we can already infer the most important features of the single-impurity problem.
First, from Eq.~(\ref{easy}), we see that the symmetries of the Hamiltonian allow for an EAA term that is proportional to $S_y^2$. Second, as $B(x_0, i \omega)$ given in Eq.~(\ref{B2}) is a real-valued function, this EAA term is always negative and therefore tends to align the spin of a single magnetic impurity along the edge of the TSC. Third, it is important to mention that $B (x_0, i \omega)$ decays into the bulk, i.e., $B (x_0 \to + \infty, i \omega) = 0$. This is due to the invariance under inversion of the $x$ axis deep in the bulk, which forces $B(x_0, i \omega)$ to vanish for $x_0 \to +\infty$. As such, the EAA is only significant close to the TSC edge at $x_0 \lesssim \xi$.

We now proceed by studying a discretized version of $\mathcal{H}+\mathcal{H}_\mathrm{imp}$ via numerical exact diagonalization, see Appendix~\ref{app:tightbinding} for details. The (classical) impurity spin is parametrized as $\bs{S} = S\,(\sin\theta \sin\phi, \cos\theta, \sin\theta \cos\phi)$ with $\theta \in \left[0, \pi \right]$ and $\phi \in \left[0, 2\pi \right)$ being the polar and azimuthal angles with respect to the $y$ axis, respectively. The ground state energy of the system can be expressed as 
\begin{align}
	E_{\mathrm{tot}} = E_0 + E_{\mathrm{imp}},
\end{align}
where $E_0$ is the energy of the clean system (i.e., in the absence of the magnetic impurity) and $E_{\mathrm{imp}}$ embodies the impurity-induced contribution. We calculate these energies by exact numerical diagonalization and display our results in Fig.~\ref{fig:SingleImpurityTotalEnergy}. In Fig.~\ref{fig:SingleImpurityTotalEnergy}(a), we plot the part of $E_{\mathrm{imp}}$ that varies as a function of $\theta$ for different distances from the edge, while we have verified that $E_{\mathrm{imp}}$ is independent of $\phi$. We find that the impurity-induced contribution takes the form
\begin{align}
	E_{\mathrm{imp}}(\theta) = C_1 + C_2 \cos^2 \theta,
	\label{eq:SingleImpurityFit}
\end{align}
with an isotropic energy shift $C_1$ and an anisotropic contribution $C_2$ depending on the orientation of the impurity spin. Figure~\ref{fig:SingleImpurityTotalEnergy}(b) displays $C_2 = E_{\mathrm{imp}}(0) - E_{\mathrm{imp}}(\pi/2)$ as a function of the distance from the edge. We find that $C_2$ is always negative, meaning that the impurity spin is favored to align along the edge ($\theta = 0$ or $\theta = \pi$). Furthermore, $C_2$ decays exponentially with the distance from the edge on a characteristic length scale $\xi$ and vanishes deep in the bulk of the system. All of these features are fully consistent with the analytical result presented in Eqs.~(\ref{easy}) and (\ref{B2}).

To make an explicit connection to the analytical result, we also evaluate Eq.~(\ref{B2}) by plugging in the numerically obtained energies and wave functions of a discretized semi-infinite system with a finite width along the $x$ direction and $k_y$ as a good quantum number. By identifying $S_y=S\cos\theta$, the value of $C_2$ can readily be obtained from Eq.~(\ref{easy}), see again Fig.~\ref{fig:SingleImpurityTotalEnergy}(b). This does indeed perfectly reproduce the result obtained via full exact diagonalization of the finite 2D system. The dependence of $C_2$ on the exchange coupling constant $J$ is analyzed numerically in Appendix~\ref{app:jdependence}, where we confirm $C_2 \propto J^2$ at $J \lesssim 1/(m S)$ as expected from perturbation theory. Contrary to that, in the strong coupling regime $J \gg 1/(m S)$, the dependence on $J$ is no longer quadratic. The parameters in Figs.~\ref{fig:SingleImpurityTotalEnergy}(a) and (b) were chosen deep in the perturbative regime in order to allow for a direct comparison between the exact numerical and the perturbative analytical results. However, we note that the magnitude of $C_2$ can be increased by about two orders of magnitude if $J$ is chosen to be larger, see again Appendix~\ref{app:jdependence}.

Last but not least, we note that our findings can be generalized to more complicated edge geometries. In this case, the single-impurity EAA term takes a more general form such that the energetically preferred configuration has the impurity spin aligned tangentially to the edge, see Fig.~\ref{fig:SingleImpurityTotalEnergy}(c).


\section{RKKY interaction}

We now analyze the interaction between two magnetic impurities through the exchange of a particle-hole pair:
\begin{align}
 E_{\mathrm{RKKY}} = J^2 \int\limits_{-\infty}^\infty \frac{d\omega}{2 \pi} {\rm Tr}\{&\sig \cdot \bm S_1 G(\bm r_1, \bm r_2, i \omega) \nonumber\\&\times\sig \cdot \bm S_2 G(\bm r_2, \bm r_1, i \omega) \} , \label{Eint}
\end{align}
where $\bm S_1$ and $\bm S_2$
are the spins of two magnetic impurities
located at $\bm r_1$ and $\bm r_2$, see Fig.~\ref{fig:SystemSketch}.
If the distance between the impurities along the edge $\ell:=|y_1 - y_2|$
is much larger than $\xi$, the contribution of the bulk states is exponentially suppressed
and only the gapless Majorana edge states 
contribute to the RKKY interaction. Evaluating Eq.~(\ref{Eint}) by taking into account only the edge states [see Eq.~(\ref{edgegap})], we find after a straightforward calculation outlined in Appendix~\ref{app:rkky}:
\begin{align}
&E_{\mathrm{RKKY}} \approx -(J m)^2 \Delta \, F(\bm r_1, \bm r_2) S_1^y S_2^y , \label{rkky} \\
&F(\bm r_1, \bm r_2) = \frac{8}{\pi}  \frac{\sin^2(k_F x_1) \sin^2(k_F x_2)}{k_F \ell} e^{-\frac{2}{\xi}(x_1 + x_2)}, \label{F}
\end{align}
which is valid in the limit $\ell \gg \xi$. Thus, the RKKY interaction is of Ising type and tends to align the impurity spins ferromagnetically along the edge~\footnote{We note that Ref.~\cite{Eriksson2015} found an AFM alignment for spin-$1/2$ impurities close to the edge of a 2D TSC. However, when revisiting the corresponding calculation (see Appendix~\ref{app:ref90}), we were not able to reproduce this result. Instead, we found that an FM alignment is preferred. As such, quantum impurities seem to behave qualitatively similar to the classical impurities discussed in our work.}. Since also the single-impurity term favors to align each individual impurity spin along the edge, the overall ground state will have both impurity spins aligned ferromagnetically along the edge. In accordance with our intuition for 1D systems, we find that $E_{RKKY}$ decays as $1/\ell$. Furthermore, we note that $E_{RKKY}$ oscillates with $x_{1,2}$ due to the spatial profile of the edge state wave functions, while there are no oscillations with $\ell$ as a direct consequence of particle-hole symmetry.

\begin{figure}[!bt]
	\centering
	\includegraphics[width=0.9\columnwidth]{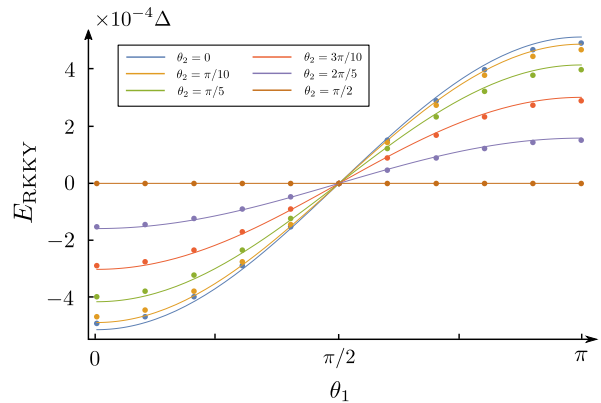}
	\caption{$E_{\mathrm{RKKY}}$ as a function of $\theta_1$ calculated numerically via exact diagonalization (dots) for different fixed $\theta_2$ (see inset) and comparison to the analytical Eq.~(\ref{rkky}) (solid lines). The separation between the impurities is $\ell=3\,\xi$ and both impurities are placed at a distance $x_1=x_2=0.2\,\xi$ from the edge. We see that, for a fixed orientation of the second impurity spin, the first impurity spin is favored to align along the edge. The total energy is minimized if both impurity spins are aligned ferromagnetically along the edge. Here, we have set $\Delta/\varepsilon_F=0.28$, $JSm=0.1$, and $\phi_1=\phi_2=0$.
}
	\label{fig:RKKY_angle}
\end{figure}

Again, we verify these results via exact diagonalization of a discretized model. The total ground state energy of the system now consists of three components: the energy of the clean system $E_0$, the impurity-induced contributions $E^{(1)}_{\mathrm{imp}}$ and $E^{(2)}_{\mathrm{imp}}$ of the individual impurities, and the RKKY exchange energy $E_{\mathrm{RKKY}}$. Thus, we can write
\begin{align}
E_{\mathrm{tot}} = E_0 + E^{(1)}_{\mathrm{imp}} +E^{(2)}_{\mathrm{imp}} + E_{\mathrm{RKKY}}.
\end{align}
Again, the two impurity spins are parametrized as $\bs{S}_i = S\,(\sin\theta_i \sin\phi_i, \cos\theta_i, \sin\theta_i \cos\phi_i)$ with $\theta_i \in \left[0, \pi \right]$ and $\phi_i \in \left[0, 2\pi \right)$ for $i\in\{1,2\}$. In Fig.~\ref{fig:RKKY_angle},  $E_{\mathrm{RKKY}}$ is calculated at a large interimpurity distance $\ell=3\,\xi$ in dependence on $\theta_1$ for different fixed $\theta_2$. Our numerical results show that indeed $E_{\mathrm{RKKY}}\propto\cos(\theta_1)$ with a prefactor that is well approximated by the analytical result given in Eqs.~(\ref{rkky}) and (\ref{F}). We have checked that the same dependence also holds for $\theta_2$ as expected by symmetry. Similarly, we have checked that the numerical curves are antisymmetric around $\theta_{1,2}=\pi/2$ in agreement with the analytical result, which is why we restrict ourselves to $\theta_2\in[0,\pi/2]$ in Fig.~\ref{fig:RKKY_angle}.

\begin{figure}[tb]
	\centering
	\includegraphics[width=0.9\columnwidth]{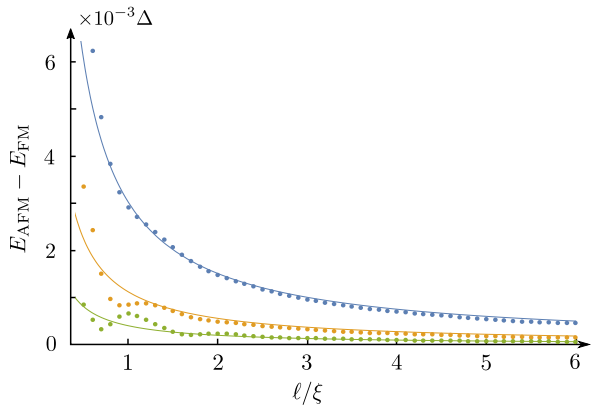}
	\caption{Energy difference $E_{\mathrm{AFM}}-E_{\mathrm{FM}}$ for two impurities oriented along the edge as a function of $\ell$ calculated numerically via exact diagonalization (dots) and analytically via Eq.~(\ref{rkky}) (solid line). The impurities are placed at a distance $x_1=x_2=0.2\,\xi$ (blue), $x_1=x_2=0.3\,\xi$ (orange), or $x_1=x_2=0.6\,\xi$ (green) from the edge. We find that $E_{\mathrm{AFM}}-E_{\mathrm{FM}}>0$, indicating that the FM configuration is energetically favorable over the AFM configuration. Furthermore, the energy difference decays
	as $1/\ell$ with increasing interimpurity distances, while it vanishes exponentially as the impurities are moved into the bulk. At small $\ell$, the exact numerical result shows additional oscillations due to contributions from the bulk states, which were neglected in the analytical treatment. The system parameters are the same as in Fig.~\ref{fig:RKKY_angle}.
}
	\label{fig:RKKY_distance}
\end{figure}

Next, we calculate the energy difference between the antiferromagnetic (AFM) and the ferromagnetic (FM) configuration for two impurities oriented along the edge as a function of $\ell$. Figure~\ref{fig:RKKY_distance} shows the results for several different distances from the edge. As expected, the energy difference 
is positive---indicating that the FM configuration is energetically favorable over the AFM configuration---and decays
as $1/\ell$ with increasing interimpurity distances. In addition to the numerical result, we also display the analytical energy difference obtained from Eqs.~(\ref{rkky}) and (\ref{F}), see Fig.~\ref{fig:RKKY_distance}. Indeed, in the limit of large $\ell\gg\xi$, the analytical expression approximates the numerical result very well. For small impurity separations, however, we see additional oscillations in the energy difference that become more pronounced as $x_{1,2}$ increase. These can be attributed to bulk contributions, which were neglected in the analytical treatment. As the impurities move away from the edge, the edge-state contribution to the RKKY interaction becomes exponentially suppressed and bulk contributions are no longer negligible. We also note that, while the parameters in Figs.~\ref{fig:RKKY_angle} and \ref{fig:RKKY_distance} were chosen such that we stay deep in the perturbative regime $JSm\ll 1$, $E_{RKKY}$ can take on significantly larger values if $J$ is increased. Indeed, $E_{RKKY}$ will grow approximately quadratically with $J$ up to $JSm\lesssim 1$.

We have checked numerically that the asymptotic $1/\ell$ decay of the RKKY interaction close to the edge also persists in the presence of potential disorder as long as the bulk gap remains open. Contrary to this, accidental edge states in topologically trivial materials (caused by, e.g., local imperfections of the edge) will not lead to such a stable $1/\ell$ decay nor to the predicted spin anisotropies. Indeed, trivial boundary states will generally be localized along certain parts of the boundary instead of propagating all the way around it in a topologically protected way. As such, they can only give an exponentially decaying contribution to the RKKY interaction.

\section{Conclusion}

We have studied classical magnetic impurities at the edge of a helical TSC both analytically and numerically.
For an isolated magnetic impurity, we have found that a strong easy-axis anisotropy tends to align the impurity spin along the edge. Furthermore, we have shown that the RKKY interaction between two magnetic impurities placed close to the edge results in a ferromagnetic alignment of both impurity spins along the edge.

Our results indicate that spectroscopy of dilute magnetic impurities can serve as a powerful tool to experimentally distinguish between topologically trivial and nontrivial materials by probing the magnetic anisotropy induced by the helical edge states. When observed together with additional signatures of bulk superconductivity, these magnetic anisotropies strongly indicate a helical TSC state. As a simple experimental check, one should be able to observe that the magnetic anisotropies---together with the bulk gap and all other features of the helical TSC state---disappear when the temperature is raised above the critical temperature of the superconductor. We also note that while we focused on a simple toy model for a 2D $p$-wave TSC, our analysis can easily be adapted to more elaborate models of helical TSCs~\cite{Fu2008,Sato2009,Liu2011,Deng2012,Nakosai2012,Zhang2013,Wang2014,Parhizgar2017,Kashigawa2019,Volpez2019,Zhang2021}.

\acknowledgments
We would like to acknowledge fruitful discussions with Flavio Ronetti and Alex Mook. This work was supported by the Georg H. Endress Foundation, the Swiss National Science Foundation and NCCR QSIT. This project received funding from the European Union’s Horizon 2020 research and innovation program (ERC Starting Grant, grant agreement No 757725).

\begin{widetext}
\appendix

\section{Exact numerical diagonalization}
\label{app:tightbinding}

For our numerical calculations, we describe the TSC by the following tight-binding Hamiltonian defined on a square lattice:
\begin{align}
\bar{H} = \sum\limits_{n,m} \{ \Psi_{n,m}^\dag \left( 4t - \mu \right) \tau_z \Psi_{n,m}
- \left[ \Psi_{n,m}^\dag \left(t \tau_z + i \Delta_0\sigma_y \tau_x\right)  \Psi_{n+1,m}  
+\Psi_{n,m}^\dag \left(t \tau_z - i \Delta_0\sigma_x \tau_x \right)   \Psi_{n,m+1} + \mathrm{H.c.}\right]\},
\label{eq:HamiltonianTS}
\end{align}
written in the Nambu basis $\Psi_{n,m} = (c_{n,m,\uparrow}$, $c_{n,m,\downarrow}$, $c_{n,m,\downarrow}^\dagger$, $-c_{n,m,\uparrow}^\dagger)^T$, where $c_{n,m,\sigma}^{(\dagger)}$ annihilates (creates) an electron with spin $\sigma \in\{ \uparrow, \downarrow\}$ at the lattice site $(n,m)$. The hopping amplitude, the $p$-wave superconducting pairing amplitude, and the chemical potential are denoted $t$, $\Delta_0$, and $\mu$, respectively. The Pauli matrices $\bs{\sigma} = (\sigma_x, \sigma_y, \sigma_z)$ and $\bs{\tau} = (\tau_x, \tau_y, \tau_z)$ act in spin and particle-hole space, respectively. 
The Hamiltonian $\bar{H}$ describes a helical topological superconductor for $\Delta_0 \neq 0$ and $\mu/t \in (0, 4) \cup (4, 8)$ \cite{BernevigHughes2013}. The continuum limit of this Hamiltonian coincides with Eq.~(\ref{ham}) of the main text upon identifying $\mu=k_F^2/(2m)$, $t=1/(2ma^2)$, and $\Delta_0=\alpha/(2a)$, where $a$ denotes the lattice constant. 

Next, magnetic impurities are placed at sites $(n_i,m_i)$ close to the edge of the TSC. They are modeled by the Hamiltonian
\begin{align}
\bar{H}_{\mathrm{imp}} = \bar{J}\,\sum_i\Psi_{n_i,m_i}^\dag \bs{\sigma}\cdot\bs{s}_i\,\Psi_{n_i,m_i},
\label{eq:HamiltonianImp}
\end{align} 
where $\bs{s}_i=\bs{S}_i/S$ is a unit vector pointing along the direction of the impurity spin at site $i$ and the exchange coupling $\bar{J}$ is related to the continuum parameters as $\bar{J}=JS/a^2$. We recall from the main text that the direction of the classical impurity spins is parametrized as $\bs{s}_i = (\sin\theta_i \sin\phi_i, \cos\theta_i, \sin\theta_i \cos\phi_i)$ with $\theta_i \in \left[0, \pi \right]$ and $\phi_i \in \left[0, 2\pi \right)$ being the polar and azimuthal angles with respect to the $y$ axis, respectively.

\section{Easy-axis anisotropy}
\label{app:eaa}

In this Appendix, we derive Eq.~(\ref{easy}) in the main text.
The Matsubara Green function $G(\bm r, \bm r', i\omega)$
for the Hamiltonian Eq.~(\ref{ham})
takes the following form:
\begin{eqnarray}
&& G(\bm r, \bm r', i\omega) = \sum_{\eta, k_y, n} \frac{\Psi_{\eta, k_y, n}(x) \Psi^\dagger_{\eta, k_y, n} (x') e^{i k_y (y - y')}}{i \omega - \varepsilon_{\eta, k_y, n}} , \label{green}
\end{eqnarray}
where $\varepsilon_{\eta, k_y, n}$ is the eigenenergy of the quantum state $\Psi_{\eta, k_y, n} (x)$
[see Eq.~(\ref{ky})] and we define $\bm r = (x, y)$, $\bm r' = (x', y')$ with
$x,x' > 0$.
The Green function depends only on the difference $y - y'$ due to the translational invariance along the $y$ direction.
The block-diagonal symmetry of the Hamiltonian, $[\mathcal{H}, \sigma_z \tau_z] = 0$, results in the corresponding block-diagonal 
structure of the Green function:
\begin{eqnarray}
&& \left[G(\bm r, \bm r', i \omega),\sigma_z \tau_z \right] = 0 . \label{Gblock}
\end{eqnarray}
In order to derive Eq.~(\ref{easy}),
we need the Green function taken at the same point $\bm r = \bm r' = \bm r_0 = (x_0, y_0)$, $x_0 > 0$:
\begin{eqnarray}
&& G(x_0, i \omega) \equiv G(\bm r_0, \bm r_0, i \omega) = \sum_{\eta, k_y, n} \frac{\Psi_{\eta, k_y, n}(x_0) \Psi^\dagger_{\eta, k_y, n} (x_0)}{i \omega - \varepsilon_{\eta, k_y, n}} . \label{Gxw0}
\end{eqnarray}
The new short-hand notation $G(x_0, i\omega)$ shows that $G(\bm r_0, \bm r_0, i \omega)$
depends only on $x_0$ and not on $y_0$. Using the symmetries of $\mathcal{H}$, we will now find the matrix
decomposition of $G(x_0, i \omega)$, which results in the EAA in Eq.~(\ref{easy}).

First, the Hamiltonian $\mathcal{H}$ is symmetric under the inversion of the $y$ axis:
\begin{eqnarray}
&& \sigma_y \mathcal{H}(-k_y) \sigma_y = \mathcal{H}(k_y) . \label{reflection}
\end{eqnarray}
This symmetry results in the following constraint for the Green function $G(x_0, i\omega)$:
\begin{eqnarray}
&& \left[\sigma_y, G(x_0, i \omega)\right] = 0.
\label{sym}
\end{eqnarray}
Second, $\mathcal{H}$ is also time-reversal symmetric:
\begin{eqnarray}
&& \sigma_y \mathcal{H}^*(-k_y) \sigma_y = \mathcal{H}(k_y) ,
\end{eqnarray}
Combined with the reflection symmetry Eq.~(\ref{reflection}), this results in the following constraint:
\begin{eqnarray}
&& \mathcal{H}^*(k_y) = \mathcal{H}(k_y) . \label{tr}
\end{eqnarray}
Here $^*$ stands for the complex conjugation. Thus, the spinors $\Psi_{\eta, k_y, n}(x)$
can be chosen real-valued:
\begin{eqnarray}
&& \Psi_{\eta, k_y, n}^*(x) = \Psi_{\eta, k_y, n}(x) . \label{real}
\end{eqnarray}
This results in the following symmetry of the Green function:
\begin{eqnarray}
&& G^*(x_0, -i \omega) = G(x_0, i \omega) . \label{Greal}
\end{eqnarray}
In fact the real-valuedness of the spinors, the reflection symmetry Eq.~(\ref{sym}), and the block-diagonal structure
of the Green function Eq.~(\ref{Gblock}) result in the following matrix decomposition 
of $G(x_0, i \omega)$:
\begin{eqnarray}
&& G(x_0, i\omega) = i A_1(x_0, i \omega) + A_2 (x_0, i \omega) \tau_z + i B (x_0, i \omega) \sigma_y \tau_y . \label{Gmat}
\end{eqnarray}
Here, $A_{1,2} (x, i \omega)$ and $B(x, i \omega)$ are complex functions such that
$A_1^*(x, -i \omega) = - A_1 (x, i \omega)$,
$A_2^*(x, -i \omega) = A_2 (x, i \omega)$, and
$B^*(x, - i \omega) = - B (x, i \omega)$.
We note that the absence of a $\sigma_y \tau_x$ term in $G(x_0, i \omega)$ is somewhat non-trivial here.
It becomes clear from the matrix structure of the projectors
$\Psi_{\eta, k_y, n}(x_0) \Psi^\dagger_{\eta, k_y, n}(x_0)$,
where $\Psi_{\eta, k_y, n}(x_0)$ is real-valued and also an eigenvector of $\sigma_z \tau_z$.

The last symmetry in play is the intrinsic particle-hole symmetry of the superconducting system. In combination with time-reversal symmetry, this gives rise to the chiral symmetry
\begin{eqnarray}
&& \tau_y \mathcal{H}(k_y) \tau_y = -\mathcal{H}(k_y), \label{ph}\\
&& G(x_0, i \omega) = - \tau_y G(x_0, -i \omega) \tau_y . \label{Gph}
\end{eqnarray}
Together with Eq.~(\ref{Greal}), Eq.~(\ref{Gph})
obliges the functions $A_{1,2} (x_0, i \omega)$ and $B(x_0, i \omega)$ 
to be real-valued. Moreover, $A_1 (x_0, i \omega)$ and $B(x_0, i \omega)$ are odd functions of $\omega$, whereas
$A_2 (x_0, i \omega)$ is an even function of $\omega$.

At this point the matrix decomposition of $G(x_0, i \omega)$, see Eq.~(\ref{Gmat}),
becomes especially handy.
First of all, it is clear that there is no 
local spin magnetization because ${\rm Tr}\{G(x_0, i \omega) \sig\} = 0$
in accordance with time-reversal symmetry.
As the first two terms in Eq.~(\ref{Gmat}) 
commute with $\sig \cdot \bm S$, i.e.,
\begin{eqnarray}
&& [i A_1 (x_0, i \omega) + A_2 (x_0, i \omega) \tau_z, \sig \cdot \bm S] = 0 ,
\end{eqnarray}
we can significantly simplify the spin trace in Eq.~(6):
\begin{eqnarray}
&& {\rm Tr} \left\{\left[\sig \cdot \bm S \, G(x_0, i \omega) \right]^2 \right\} = - B^2(x_0, i \omega) {\rm Tr} \left\{\left(\sig \cdot \bm S \sigma_y \tau_y\right)^2
\right\} + 4 S^2 \left(A_2^2 (x_0, i \omega) - A_1^2 (x_0, i \omega)\right) , \\
&& (\sig \cdot \bm S \sigma_y \tau_y)^2 = \left(S_y + i S_x \sigma_z - i S_z \sigma_x \right)^2  = S_y^2 - S_x^2 - S_z^2 = 2 S_y^2 - S^2 .
\end{eqnarray}
This directly leads us to Eq.~(\ref{easy}).	

Following the matrix decomposition Eq.~(\ref{Gmat}), 
we can represent $B (x_0, i \omega)$
in the following form:
\begin{eqnarray}
&& B (x_0, i \omega) = - \frac{i}{4} {\rm Tr} \left\{\sigma_y \tau_y G(x_0, i \omega) \right\} . \label{B0}
\end{eqnarray}
We note that $B (x_0, i \omega)$ is an odd function of $\omega$ due to chiral symmetry, see Eq.~(\ref{ph}).
Thus, we can simplify Eq.~(\ref{B0}) by taking the antisymmetric part of the Green function in Eq.~(\ref{Gxw0}),
which results in the following representation:
\begin{eqnarray}
&& B (x_0, i \omega) = - \frac{\omega}{4} \sum\limits_{\eta, k_y, n} \frac{\Psi^\dagger_{\eta, k_y, n}(x_0) \sigma_y \tau_y \Psi_{\eta, k_y, n} (x_0)}{\omega^2 + \varepsilon_{\eta, k_y, n}^2} . \label{B2a}
\end{eqnarray}
From the chiral symmetry, it is also clear that the contribution of the states with negative energies is the 
same as the contribution coming from the positive energies, which results in Eq.~(\ref{B2}) in the main text.

\section{Local density of states in the vicinity of an impurity}
\label{app:ldos}

To begin with, let us calculate the local density of states for the unperturbed TSC. The retarded Green function of the unperturbed system taken at $\bm r = \bm r'$ is the following:
\begin{eqnarray}
&& G_R^{(0)}(\bm r, \bm r, \omega) = \sum_{\eta, k_y, n} \frac{\Psi_{\eta, k_y, n}(x) \Psi^\dagger_{\eta, k_y, n} (x)}{\omega - \varepsilon_{\eta, k_y, n} + i0^+} , \label{Gxw}
\end{eqnarray}
where $\eta = \pm 1$ is the eigenvalue of $\sigma_z \tau_z$,
$k_y$ is the momentum along $y$,
$x > 0$ is the distance from the edge, and
$n$ corresponds to all other quantum numbers in the system.
The local density of states is described by the following operator:
\begin{equation}
\rho^{(0)}(\bm r, \bm r, \omega) = -\frac{1}{\pi} \textrm{Im} \left\{G_R^{(0)}(\bm r, \bm r, \omega)
\right\} = \sum_{\eta, k_y, n} \Psi_{\eta, k_y, n}(x) \Psi^\dagger_{\eta, k_y, n} (x) \delta \left(\omega - \varepsilon_{\eta, k_y, n}\right) .
\end{equation} 
If $|\omega| < \Delta$,
then only the in-gap edge states contribute to the local density of states.
This allows us to consider only the edge states without the bulk contribution.
The edge state wave function 
is given by Eq.~(\ref{edgegap}):
\begin{eqnarray}
&& \Psi_{\eta, k_y}(x) = \phi(k_y, x) \chi_\eta ,
\end{eqnarray}
where $\chi_{\eta}$ is the Majorana spinor and $\phi(k_y, x)$ is the Majorana wave function normalized to unity:
\begin{eqnarray}
&& \phi(k_y, x) = 2 \sqrt{\frac{k_F^2 - k_y^2}{\xi}} \frac{\sin \left(\kappa x\right)}{\kappa} e^{-\frac{x}{\xi}}.
\end{eqnarray}
%
The spectrum of the edge states is $\varepsilon_{\eta, k_y} = -\eta \alpha k_y$, see the main text.
With this, we find the edge state contribution to the unperturbed local density of states:
\begin{equation}
\rho^{(0)}_e(\bm r, \bm r, \omega) = \int\limits_{-k_F}^{k_F} \frac{d k_y}{2 \pi} \phi^2(k_y, x) \delta(\omega - \alpha k_y) \mathcal{P} = \frac{\phi^2\left(\frac{\omega}{\alpha}, x\right)}{2 \pi \alpha} \mathcal{P} \approx \frac{2 m}{\pi} e^{-\frac{2 x}{\xi}} \mathcal{P} \sin^2 \left(\frac{x}{\alpha}\sqrt{\Delta^2 - \omega^2}\right), \label{rho0}
\end{equation}
where $\mathcal{P} = (1 + \sigma_y \tau_y)/2$ is the projector onto the Majorana edge mode subspace $\sigma_y \tau_y = +1$.
Here $|\omega| < \Delta \approx \alpha k_F$ and we approximated the ratio $(\alpha^2 k_F^2 - \omega^2)/[\alpha^2 k_F^2 - \omega^2 - (m \alpha^2)^2] \approx 1$
if $\Delta^2 - \omega^2 \gg (m \alpha^2)^2$.

The analyticity of $G_R^{(0)}(\bm r, \bm r, \omega)$ in the upper half-plane ${\rm Im}(\omega) > 0$
allows us to restore the edge state contribution to the retarded Green function:
\begin{equation}
G_{R,e}^{(0)}(\bm r, \bm r, \omega) = i m e^{-\frac{2 x}{\xi}} \left(\exp\left(-\sqrt{(\omega + i 0^+)^2 -\Delta^2} \frac{2 x}{\alpha}\right)-1\right) \mathcal{P} . \label{GR0}
\end{equation}
Notice that this function is indeed analytic in the upper half-plane ${\rm Im}(\omega) > 0$,
satisfies the boundary condition $G_{R,e}^{(0)}(\bm r, \bm r, \omega) = 0$ at $x = 0$,
and the imaginary part yields the local density of states given by Eq.~(\ref{rho0}). The bulk contribution, on the other hand, can be estimated by approximating the bulk density of states 
by the value $\rho_b(\bm r, \bm r, \omega) \approx m/\pi$ for $|\omega| > \Delta$ and zero otherwise:
\begin{equation}
G_b(\bm r, \bm r, \omega) = \int\limits_{-\infty}^\infty \frac{\rho_b(\bm r, \bm r, \Omega) \, d\Omega}{\omega - \Omega + i0^+} \approx -\frac{2 m \omega}{\pi} \int\limits_\Delta^\infty \frac{d\Omega}{\Omega^2 - (\omega + i 0^+)^2} = - \frac{m}{\pi} \ln \left(\frac{\Delta + \omega+ i0^+}{\Delta - \omega - i0^+}\right) .
\end{equation}
We are interested in the local density of states at small frequencies $|\omega| \ll \Delta$, where the edge contribution Eq.~(\ref{GR0}) dominates over the bulk state contribution.
This corresponds to the condition
\begin{eqnarray}
&& |\omega| \ll \Delta e^{-\frac{2 x}{\xi}} . \label{omcond}
\end{eqnarray}
Equation~(\ref{GR0}) can be further simplified at $|\omega| \ll \Delta$ by using $\sqrt{(\omega + i0^+)^2 - \Delta^2} \approx i \Delta {\rm sgn}(\omega)$:
\begin{eqnarray}
&& G^{(0)}_{R,e}(\bm r, \bm r, \omega) \approx i m e^{-\frac{2 x}{\xi}} \left(e^{-2 i k_F x \, {\rm sgn}(\omega)} - 1\right) \mathcal{P} . \label{GRedge}
\end{eqnarray}

From Eq.~(\ref{rho0}) we immediately see that there is no magnetization in the unperturbed system:
\begin{eqnarray}
&& \mbox{\boldmath{$\rho$}}_s^{(0)}(\bm r, \bm r, \omega) \equiv {\rm Tr}\left\{\mbox{\boldmath{$\sigma$}} \rho^{(0)}(\bm r, \bm r, \omega) \right\} = 0 ,
\end{eqnarray}
where the index $s$ in $\mbox{\boldmath{$\rho$}}_s^{(0)}(\bm r, \bm r, \omega)$ stands for spin.
The local charge density of states $\rho_c^{(0)}(\bm r, \bm r, \omega)$ at $|\omega| \ll \Delta$ of the unperturbed system follows from Eq.~(\ref{GRedge}):
\begin{equation}
\rho_c^{(0)}(\bm r, \bm r, \omega) \equiv {\rm Tr} \left\{ \rho^{(0)}(\bm r, \bm r, \omega)
\right\} = \frac{\phi^2\left(\frac{\omega}{\alpha}, x\right)}{\pi \alpha}  \approx \frac{4 m}{\pi} e^{-\frac{2 x}{\xi}} \sin^2\left(k_F x\right) . \label{rhoc0}
\end{equation}

Now, we calculate the local density of states in the vicinity of a magnetic impurity.
Adding the magnetic impurity at $\bm r_0 = (x_0, y_0)$ [see Eq.~(\ref{spinspin}) in the main text]
perturbs the electron Green function in the vicinity of the impurity:
\begin{equation}
G(\bm r, \bm r', i \omega) = G^{(0)}(\bm r, \bm r', i \omega) + J G^{(0)}(\bm r, \bm r_0, i \omega) \sig \cdot \bm S G(\bm r_0, \bm r', i \omega) . \label{Gexact}
\end{equation}
%
Here we use the Matsubara formalism. The retarded Green function can be obtained with the help of the analytical continuation 
$i \omega \to \omega + i 0^+$.
First, let us plug in $\bm r = \bm r_0$ and solve for $G(\bm r_0, \bm r', i \omega)$:
\begin{equation}
G(\bm r_0, \bm r', i \omega)  = \left(1 - J G^{(0)}(\bm r_0, \bm r_0, i \omega) \sig \cdot \bm S\right)^{-1} G^{(0)}(\bm r_0, \bm r', i \omega) .
\end{equation}
Substituting this back into Eq.~(\ref{Gexact}) and performing the analytical continuation $i \omega \to \omega+i0^+$,
we find the exact retarded Green function of the TSC with a single magnetic impurity:
\begin{equation}
G_R (\bm r, \bm r', \omega) = G_R^{(0)}(\bm r, \bm r', \omega) + J G_R^{(0)}(\bm r, \bm r_0, \omega) \sig \cdot \bm S  \left(1 - J G_R^{(0)}(\bm r_0, \bm r_0, \omega) \sig \cdot \bm S\right)^{-1} G_R^{(0)}(\bm r_0, \bm r', \omega) . \label{GRrrp}
\end{equation}

First, let us check that there are no poles induced by the impurity at $|\omega| \ll \Delta$.
For this, we notice that $G_{R,e}^{(0)}(\bm r_0, \bm r_0, \omega) \propto \mathcal{P}$, see Eq.~(\ref{GRedge}).
Then, we notice that $\mathcal{P} \sig \cdot \bm S \mathcal{P} = S_y \sigma_y \mathcal{P}$.
This is enough to simplify the following operator:
\begin{eqnarray}
&& (1 - z \mathcal{P} \sig \cdot \bm S)^{-1}  = 1 + \frac{1 + z S_y \sigma_y}{1 - z^2 S_y^2} z \mathcal{P} \sig \cdot \bm S , \label{invert}
\end{eqnarray}
where $z$ is a complex number.
There is a pole  if $z^2 S_y^2 = 1$.
In our case $z$ is the following number:
\begin{equation}
z = \frac{J}{2} {\rm Tr}\left\{G_{R,e}^{(0)}(\bm r_0, \bm r_0, \omega) \right\}  = {\rm sgn}(\omega) 2 J m e^{-\frac{2 x_0}{\xi}}\sin\left(k_F x_0\right) e^{-i {\rm sgn}(\omega) k_F x_0} . \label{z}
\end{equation}
Notice that the unperturbed retarded Green function can be conveniently written through $z$:
\begin{eqnarray}
&& G_{R,e}^{(0)}(\bm r_0, \bm r_0, \omega) = \frac{z}{J} \mathcal{P} . \label{Gz}
\end{eqnarray}
The pole condition can be represented as
\begin{eqnarray}
&& (2 J m S_y)^2 e^{-\frac{4 x_0}{\xi}} \sin^2\left(k_F x_0\right)  = e^{2i {\rm sgn}(\omega) k_F x_0} . \label{pole}
\end{eqnarray}
As before, $|\omega| \ll \Delta e^{-\frac{2 x_0}{\xi}}$, see Eq.~(\ref{omcond}).
We see that the left-hand side of Eq.~(\ref{pole}) is a positive real number.
The right-hand side is a positive real number only if $k_F x_0 = \pi n$, $n$ being an integer.
However, $\sin(k_F x_0) = \sin(\pi n) = 0$, which means that Eq.~(\ref{pole}) does not have any solutions.
This confirms our numerical simulations, which did not identify any bound states near the magnetic impurity. Notice that bound states are also impossible at any $|\omega|< \Delta$ because the bulk contribution to $G_R^{(0)}(\bm r_0, \bm r_0, \omega)$ is real-valued at any $|\omega| < \Delta$,
while the edge contribution always contains an imaginary part.
However, this argument no longer applies in the bulk where $x_0 \gg \xi$ as the edge contribution vanishes exponentially.

Equation~(\ref{GRrrp}) becomes especially simple
if we choose $\bm r = \bm r' = \bm r_0$:
\begin{equation}
G_R (\bm r_0, \bm r_0, \omega) = \left(1 - J G_R^{(0)}(\bm r_0, \bm r_0, \omega) \sig \cdot \bm S\right)^{-1} G_R^{(0)}(\bm r_0, \bm r_0, \omega) .
\end{equation}
Using Eq.~(\ref{invert}) and the representation given in Eq.~(\ref{Gz}), we can find the correction to the Green function:
\begin{equation}
\delta G_R(\bm r_0, \bm r_0, \omega) \equiv G_R(\bm r_0, \bm r_0, \omega) - G_R^{(0)}(\bm r_0, \bm r_0, \omega) = \frac{1 + z S_y \sigma_y}{1 - z^2 S_y^2} \frac{z^2}{J} \mathcal{P} \sig \cdot \bm S \mathcal{P} .
\end{equation}
Using that $\mathcal{P} \boldsymbol{\sigma}  \cdot \bm S \mathcal{P} = S_y \sigma_y \mathcal{P}$, we obtain:
\begin{eqnarray}
&& G_R(\bm r_0, \bm r_0, \omega) = \frac{z}{J} \frac{1 + z S_y \sigma_y}{1 - z^2 S_y^2} \mathcal{P} . \label{GRr0}
\end{eqnarray}
The local charge density of states at $|\omega|\ll \Delta e^{-\frac{2 x_0}{\xi}}$
is then the following:
\begin{eqnarray}
&& \rho_c (\bm r_0, \bm r_0, \omega) = \frac{\left(1 + L\right) \rho_c^{(0)}(\bm r_0, \bm r_0, \omega)}{1 - 2 L \cos(2 k_F x_0) + L^2} \quad \mbox{ with }\quad L = \left[2 J S_y m e^{-2 x_0/\xi} \sin(k_F x_0)\right]^2 ,\label{rhoc}
\end{eqnarray}
where $\rho_c^{(0)}(\bm r_0, \bm r_0, \omega)$ is the unperturbed density of states, see Eq.~(\ref{rhoc0}).
In the perturbative regime $J S m \ll 1$, we have $L \ll 1$.
Thus, we see that the density of states is only slightly altered at the impurity site and can be correctly accounted for by perturbation theory.
The spin density of states at the impurity site is affected even less, see Eq.~(\ref{GRr0}).
The same logic can be applied for $G_R(\bm r, \bm r, \omega)$ at other sites $\bm r \ne \bm r_0$ in the vicinity of the impurity. 
Therefore, we confirm that the perturbation theory yields correct qualitative results for magnetic impurities placed at the edge of a topological superconductor when $J S m \ll 1$.

It is worth noting that there is no pole in Eq.~(\ref{rhoc}) even if $L = 1$.
Indeed, in order to have a pole we require $\cos(2 k_F x_0) = 1$, i.e., $\sin(k_F x_0) = 0$ and therefore
$L \propto \sin^2(k_F x_0) = 0$.
If the impurity is placed at the maximum of $\rho_c^{(0)}(\bm r_0, \bm r_0, \omega)$,
then $\sin^2(k_F x_0) = 1$ and $\cos(2 k_F x_0) = -1$.
The local charge density of states at the impurity position is therefore scaled down 
by a factor $\rho_c/\rho_c^{(0)} = 1/(L + 1)$, see Eq.~(\ref{rhoc}).
In general, the ratio $\rho_c/\rho_c^{(0)}$ can be larger or smaller than unity, depending on the impurity position.

\begin{figure*}[bt]
	\centering
	\includegraphics[width=0.9\columnwidth]{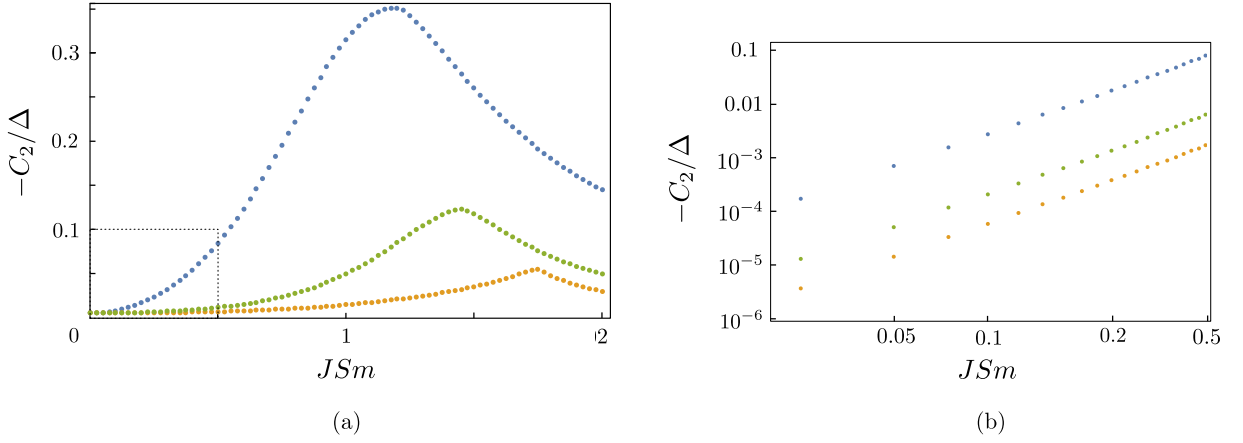}
	\caption{(a) Numerically calculated $C_2$ as a function of the exchange coupling constant $J$ for three different distances from the edge $x_0=0.2\,\xi$ (blue), $x_0=0.4\,\xi$ (orange), and $x_0=0.6\,\xi$ (green). (b) Dashed rectangle in (a) in the log-log scale. We see that the perturbation theory, which predicts an approximately quadratic dependence $C_2 \propto J^2$, is valid up to $J \lesssim 1/(m S)$. The other parameters are the same as in Fig.~2.
	}
	\label{fig:SingleImpurityJscaling}
\end{figure*}

\section{Dependence on the exchange coupling constant $J$}
\label{app:jdependence}

In order to set the limits on the perturbation theory, we present numerical results for the coefficient $C_2$ 
as a function of the exchange coupling constant $J$ for three distances from the edge, see Fig.~\ref{fig:SingleImpurityJscaling}. As expected, we find that $C_2\propto J^2$ for small $J \lesssim 1/(m S)$. 
Contrary to that, we find a qualitatively different behavior for large $J \gg 1/(m S)$, where $C_2$ decreases.
In this case, the perturbation theory fails together with the mean-field approach to describe the TSC due to a strong local renormalization of the superconducting order parameter close to the magnetic impurity.

\section{The RKKY interaction close to the edge}
\label{app:rkky}

Here we derive Eq.~(\ref{rkky}) from Eq.~(\ref{Eint}).
As we consider the limit $\ell=|y_1 - y_2| \gg \xi$,
the contribution of the bulk states is exponentially suppressed,
so we can focus on the edge state contribution alone.
Using the exact wave functions for the edge states, see Eq.~(\ref{edgegap}),
we find the Green function asymptotics at 
$\ell \gg \xi$:
\begin{eqnarray}
&& G(\bm r_1, \bm r_2, i \omega)  \approx \sum\limits_{\eta} \int\limits_{-k_F}^{k_F} \frac{d k_y}{2 \pi} \frac{\Psi_{\eta, k_y}(x_1) \Psi^\dagger_{\eta, k_y}(x_2) e^{i k_y (y_1-y_2)}}{i \omega + \eta \alpha k_y} . 
\label{Gedge}
\end{eqnarray}
As the integral over $k_y$ converges 
on the scale $k_y \sim 1/\ell \ll m \alpha \ll k_F$,
we can safely neglect the $k_y$ dependence in 
the spinors Eq.~(\ref{edgegap}):
\begin{eqnarray}
&& \Psi_{\eta, k_y} (x) \approx \phi(x) \frac{1}{\sqrt{2}} {u_\eta \choose -\eta u_{-\eta}}, \label{k0}\\
&& \phi(x) = 2 \sqrt{m \alpha} \sin (k_F x) e^{- \frac{x}{\xi}} . \label{phi}
\end{eqnarray}
Here $\phi(x)$ is normalized by unity.
Performing the summation over $\eta$ in Eq.~(\ref{Gedge})
and using the approximation Eq.~(\ref{k0}),
we find:
\begin{eqnarray}
&& G(\bm r_1, \bm r_2, i \omega) \approx \phi(x_1) \phi(x_2) \int\limits_{-\infty}^\infty \frac{dk_y}{2 \pi} \frac{e^{i k_y (y_1-y_2)}}{i \omega + \alpha k_y \sigma_z \tau_z} \mathcal{P} ,  \label{Gky}\\
&& \mathcal{P} = \frac{I + \sigma_y \tau_y}{2} . \label{P}
\end{eqnarray}
Here we introduced the projector $\mathcal{P}$
onto the subspace $\sigma_y \tau_y  = 1$.
We also extended the integration over $k_y$
to the infinite interval due to its fast convergence on the scale of $1 / \ell$.
Evaluating the integral over $k_y$ in Eq.~(\ref{Gky}),
we find the asymptotics of the Green function at $\ell \gg \xi$:
\begin{eqnarray}
&& G(\bm r_1, \bm r_2, i \omega)  \approx i \phi(x_1) \phi(x_2) \frac{\sigma_z \tau_z {\rm sgn} (y_1-y_2) - {\rm sgn}(\omega)}{2 \alpha} e^{-\frac{|\omega (y_1-y_2)|}{\alpha}} \mathcal{P} , \label{Gasympt}
\end{eqnarray}
where ${\rm sgn}(z)$ returns the sign of a real 
variable $z$.

Finally, we substitute Eq.~(\ref{Gasympt})
into Eq.~(\ref{Eint})
to find the effective interaction between two spins 
near the edge of the topological superconductor.
In order to calculate corresponding spin trace in Eq.~(\ref{Eint}),
we use two simple facts:
\begin{eqnarray}
&& \left[\mathcal{P}, \sigma_z \tau_z\right] = 0 , \\
&& \mathcal{P} (\sig \cdot \bm S) \mathcal{P} = \frac{\sigma_y + \tau_y}{2} S_y .
\end{eqnarray}
By surrounding $(\sig \cdot \bm S_1)$ 
by the projector $\mathcal{P}$
we find that the spin trace must be proportional to $S_1^y$.
Doing the same for $(\sig \cdot \bm S_2)$,
we conclude that $E_{RKKY}$
must be just proportional to $S_1^y S_2^y$.
In other words, we can just substitute 
$\sigma_y S_y$ instead of the $(\sig \cdot \bm S)$
terms in Eq.~(\ref{Eint})
to simplify the calculation.
All in all, we find the effective interaction between two spins
near the edge of a topological superconductor:
\begin{eqnarray}
E_{RKKY}&& \approx -\frac{J^2}{\alpha^2} \phi^2(x_1) \phi^2(x_2) S_1^y S_2^y \int\limits_{-\infty}^\infty \frac{d \omega}{2 \pi} e^{-\frac{2|\omega \ell|}{\alpha}} \nonumber \\
&&= -\frac{8}{\pi}(J m )^2 S_1^y S_2^y  \sin^2(k_F x_1) \sin^2(k_F x_2) e^{-\frac{2}{\xi}(x_1 + x_2)} \frac{\Delta}{k_F \ell} , 
\end{eqnarray}
where $\ell \gg \xi$.
This coincides with Eq.~(\ref{rkky}).

\section{RKKY interaction for spin-$\frac{1}{2}$ impurities}
\label{app:ref90}

In Ref.~\cite{Eriksson2015}, the RKKY interaction between two spin-$1/2$ impurities placed at the edge of a 2D TSC was calculated using an effective 1D description of the helical Majorana edge states. It was found that an antiferromagnetic alignment of the impurity spins is preferred. Here, we repeat the corresponding calculations using the model of Ref.~\cite{Eriksson2015} and find that the RKKY interaction is ferromagnetic in agreement with our results presented in the main text.

The Majorana edge modes in Ref.~\cite{Eriksson2015} are treated within the effective 1D model
\begin{eqnarray}
&& H_0 (\tau) = \frac{1}{2} \sum\limits_\nu v_\nu \int dx \, \Psi_\nu (x, \tau) (-i \partial_x) \Psi_\nu(x, \tau) , \label{Hma} 
\end{eqnarray}
where the index $\nu \in \{L, R\}$ labels left- and right-moving modes,
$v_L = - v$ and $v_R = +v > 0$ are the corresponding velocities, and
$\Psi_\nu(x, \tau) = \Psi^\dagger_\nu (x, \tau)$ 
are the Majorana field operators satisfying the following anticommutation relation:
\begin{eqnarray}
&& \left\{\Psi_\nu (x, \tau), \Psi_{\nu'}(x', \tau) \right\} = \delta_{\nu \nu'} \delta (x - x') . \label{macom}
\end{eqnarray}
The equation of motion for the Majorana field operators in imaginary time is then the following:
\begin{eqnarray}
&& \frac{\partial \Psi_\nu (x, \tau)}{\partial \tau} = \left[H_0(\tau), \Psi_\nu (x, \tau)\right] = i v_\nu \partial_x \Psi_\nu(x, \tau) , \label{motioneq}
\end{eqnarray}
where we used the anticommutation relation Eq.~(\ref{macom}) to evaluate the commutator.

We define the imaginary time Majorana Green function as follows:
\begin{eqnarray}
&& G_{\nu \nu'}(x, \tau) = - \left< \mathrm{T}\left\{\Psi_\nu (x, \tau) \Psi_{\nu'}(0, 0) \right\}\right> = - \vartheta(\tau) \left<\Psi_\nu (x, \tau) \Psi_{\nu'}(0, 0) \right> + \vartheta(-\tau) \left< \Psi_{\nu'} (0, 0) \Psi_{\nu}(x, \tau) \right> ,
\end{eqnarray}
where $\mathrm{T}$ is the time-ordering operator and $\vartheta (\tau)$ is the Heaviside step function.
Using Eq.~(\ref{motioneq}), we find the differential equation for $G_{\nu \nu'}(x, \tau)$:
\begin{eqnarray}
&& \partial_\tau G_{\nu \nu'}(x, \tau) = - \delta (\tau) \delta (x) \delta_{\nu \nu'} + i v_\nu \partial_x G_{\nu \nu'}(x, \tau) .
\end{eqnarray}
The Green function is especially simple in the frequency-momentum representation:
\begin{eqnarray}
&& G_{\nu\nu'}(k, i \omega) = \frac{\delta_{\nu \nu'}}{i \omega - v_\nu k} .
\end{eqnarray}
Taking the Fourier transform, we find the Green function in the time-coordinate representation:
\begin{eqnarray}
&& G_{\nu \nu'}(x, \tau) = - \frac{i T}{2 v} \frac{\delta_{\nu \nu'}}{\sinh\left(\pi T \left(\frac{x}{v_\nu} + i \tau\right)\right)} , \label{gxt}
\end{eqnarray}
where $T$ is the temperature. The anti-periodic condition is satisfied:
\begin{eqnarray}
&& G_{\nu \nu'}(x, \tau + 1/T) = - G_{\nu \nu'}(x, \tau) .
\end{eqnarray}
As the Green function is diagonal with respect to the chiral index, we introduce the chiral left and right components
\begin{eqnarray}
&& G_R(x, \tau) \equiv G_{RR} (x, \tau), \hspace{5pt} G_L(x, \tau) \equiv G_{LL}(x, \tau) , \label{leftright}
\end{eqnarray}
where $G_{\nu\nu'}(x, \tau)$ is given by Eq.~(\ref{gxt}). $G_R(x, \tau)$ in this Appendix should not be confused with the retarded Green function.

We use the effective exchange interaction defined in Ref.~\cite{Eriksson2015}:
\begin{eqnarray}
&& H_J (\tau) = J \left(s\left(-\frac{R}{2}, \tau\right) S_{I,1} + s\left(\frac{R}{2}, \tau\right) S_{I,2} \right) , \label{HJ}
\end{eqnarray}
where two spin impurities are located at $x = \pm R/2$, $R$ is the distance between the impurities along the edge,
$J$ is the effective exchange coupling,
and $S_{I,1}$, $S_{I, 2}$ are the projections of the impurity spins onto the so-called Ising direction, see Ref.~\cite{Eriksson2015}.
In our case, the Ising direction is along the edge.
The Majorana spin density operator $s(x, \tau)$
is defined as follows:
\begin{equation}
s(x, \tau) = i \Psi_R (x, \tau) \Psi_L (x, \tau) .
\end{equation}

The interaction contribution to the thermodynamic potential $\delta \Omega$ is given by the following statistical average:
\begin{eqnarray}
&& \delta \Omega = - T \left(\left<\mathfrak{S} \right>_c - 1\right) , \label{Smat}
\end{eqnarray}
where the index $_c$ stands for the connected diagrams only and
$\mathfrak{S}$ is the statistical $S$-matrix:
\begin{eqnarray}
&& \mathfrak{S} = \mathrm{T}\,\mathrm{exp}\left(-\int\limits_0^{1/T} H_J(\tau) \, d\tau \right) ,
\end{eqnarray}
where $H_J(\tau)$ is given by Eq.~(\ref{HJ}).
The first-order correction vanishes due to the zero net spin density in absence of the magnetic impurities:
\begin{eqnarray}
&& \left< s(x, \tau) \right> = 0 .
\end{eqnarray}
The second-order correction yields the self-interaction terms proportional to $S_{I, 1}^2$ and $S_{I, 2}^2$ 
as well as the RKKY term proportional to $S_{I, 1} S_{I, 2}$ that we are interested in:
\begin{align}
E_{RKKY}& = - T J^2 S_{I, 1} S_{I, 2} \int d\tau_1 d\tau_2 \, \left< \mathrm{T} \left\{ s\left(-\frac{R}{2}, \tau_1\right) s\left(\frac{R}{2}, \tau_2\right) \right\} \right> \nonumber \\
& = T J^2 S_{I, 1} S_{I, 2} \int d\tau_1 d\tau_2 \, G_R (-R, \tau_1 - \tau_2) G_L (R, \tau_2 - \tau_1 ) , \label{rk}
\end{align}
where $G_R(x, \tau)$ and $G_L(x, \tau)$ are introduced in Eq.~(\ref{leftright}).
The Wick theorem has been used to evaluate the 
statistical average in Eq.~(\ref{rk}).
The product of two Green functions in Eq.~(\ref{rk}) is a periodic function on $\tau$ with the period $1/T$.
This allows for the following simplification:
\begin{eqnarray}
&& E_{RKKY} = J^2 S_{I, 1} S_{I, 2} \int\limits_0^{1/T} d\tau \, G_R(-R, \tau) G_L (R, -\tau) . \label{Etau}
\end{eqnarray}
Substituting the Green functions Eq.~(\ref{gxt})
into Eq.~(\ref{Etau}),
we find:
\begin{eqnarray}
&& E_{RKKY} = -\frac{J^2 T^2}{2 v^2} S_{I,1} S_{I, 2} \int\limits_{0}^{1/T} \frac{d \tau}{\cosh\left(2 \pi T \frac{R}{v}\right) - \cos \left(2 \pi T \tau\right)} .
\end{eqnarray}
The integral over $\tau$ can be reduced to the following elementary integral:
\begin{eqnarray}
&& \int\limits_0^{2\pi} \frac{dt}{\lambda - \cos t} = \frac{2 \pi}{\sqrt{\lambda^2 - 1}}, \, \lambda > 1 .
\end{eqnarray}
This gives the final result, showing that the RKKY interaction mediated by the helical Majorana edge modes is ferromagnetic:
\begin{eqnarray}
&& E_{RKKY} = - \frac{J^2 S_{I, 1} S_{I, 2}}{4 \pi v R} \frac{2 \pi T R / v}{\sinh \left(2 \pi T R /v\right)} .
\end{eqnarray}
As such, we find that, for vanishing temperature $T=0$, the result for quantum impurities is qualitatively similar to the result for classical impurities presented in Eq.~(\ref{rkky}).

\end{widetext}
	
\bibliography{biblio_RKKY_helical_edges}

\end{document}